\documentclass[aps,prl,twocolumn,floatfix,showpacs,preprintnumbers,superscriptaddress]{revtex4}

\usepackage{graphics}
\usepackage{graphicx}
\usepackage{epsfig}

\newcommand{\beq}{\begin{equation}}
\newcommand{\eeq}{\end{equation}}
\newcommand{\bea}{\begin{eqnarray}}
\newcommand{\eea}{\end{eqnarray}}
\usepackage{dcolumn}
\usepackage{bm}

\begin{document}

\title{Electrical Detection of Spin Accumulation at a Ferromagnet-Semiconductor Interface}

\author{X. Lou}
\affiliation{School of Physics and Astronomy, University of Minnesota, Minneapolis, MN  55455}
\author{C. Adelmann}
\altaffiliation[Current address: ]{IMEC, 3000 Leuven, Belgium}
\affiliation{Department of Chemical Engineering and Materials Science, University of Minnesota, Minneapolis, MN 55455}
\author{M. Furis}
\affiliation{National High Magnetic Field Laboratory, Los Alamos National Laboratory, Los Alamos, NM  87545}
\author{S.~A.~Crooker}
\affiliation{National High Magnetic Field Laboratory, Los Alamos National Laboratory, Los Alamos, NM  87545}
\author{C.~J.~Palmstr\o m}
\affiliation{Dept.~of Chemical Engineering and Materials Science, University of Minnesota, Minneapolis, MN 55455}
\author{P.~A. Crowell}
\affiliation{School of Physics and Astronomy, University of Minnesota, Minneapolis, MN  55455}
\email[]{crowell@physics.umn.edu}

\begin{abstract}
We show that the accumulation of spin-polarized electrons at a forward-biased Schottky tunnel barrier between Fe and $n$-GaAs can be detected electrically.  The spin accumulation leads to an additional voltage drop across the barrier that is suppressed by a small transverse magnetic field, which depolarizes the spins in the semiconductor.  The dependence of the electrical accumulation signal on  magnetic field, bias current, and temperature is in good agreement with the predictions of a drift-diffusion model for spin-polarized transport.
\end{abstract}
\maketitle

The injection and detection of spin-polarized electrons in semiconductors are two of the important problems in the physics of spin transport.  Recent experiments have demonstrated that it is possible to maintain a non-equilibrium spin polarization greater than 25\% in a semiconductor by injection of electrons from a ferromagnetic metal through a tunnel barrier \cite{Hanbicki:2003,IBM:2005,Adelmann:2005}.  In the case of a Schottky barrier, spin accumulation also occurs when electrons flow from the semiconductor to the ferromagnet \cite{Epstein:2003,Stephens:2004,Crooker:2005}.   In almost all cases, however, measuring the spin polarization in the semiconductor has required the use of optical techniques.  This raises the important question of whether an electrically generated spin polarization at a ferromagnet-semiconductor interface can also be detected electrically.  The strongest test of any electrical spin detection measurement is the existence of a Hanle effect \cite{Johnson:1985,Jedema:2002}, in which spin accumulation is suppressed by precession in a transverse magnetic field.   A Hanle effect has not been observed in electrical measurements of spin accumulation in semiconductors reported previously \cite{Hammar:2002,FertGaMnAs:2003,Schmidt:2001}.  In contrast, the {\it optical} Hanle effect has served as the basis for a series of definitive measurements of spin-dependent phenomena in semiconductors\cite{OpticalOrientation,Kato:2004}.

In this Letter, we report a direct electrical transport measurement of spin accumulation at an Fe/$n$-GaAs interface in the presence of a forward-bias current.  The existence of a spin accumulation is established using the electrical Hanle effect \cite{Johnson:1985}.  The spin-dependent voltage across the interface is suppressed by precession in a magnetic field applied perpendicular to the direction of spin polarization.  The  resulting peak in the voltage at zero magnetic field has a width determined by the timescales for spin precession, relaxation, drift, and diffusion.  There is no corresponding electrical signature of spin injection under reverse bias, in agreement with a density of states argument based on tunnelling.

\begin{figure}
    \includegraphics*[width=6.0cm]{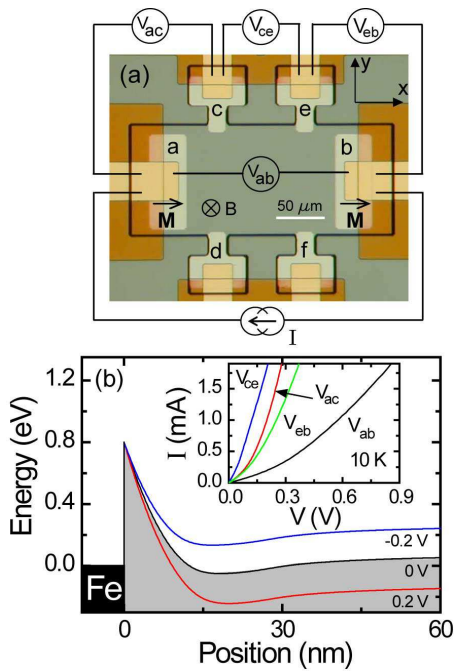}
   
    \caption{(color on-line) (a) Photomicrograph of the Fe/GaAs device. The side contacts are used for the voltage measurements indicated by the labels. (b) Band diagram near the Fe/GaAs interface under reverse bias (0.2~V), zero bias (0~V) and forward bias (-0.2~V). The inset shows the total ($V_{ab}$), drain ($V_{ac}$), source ($V_{eb}$) and channel ($V_{ce}$) voltage measurements when current flows from $a$ to $b$ at $T=10$~K.}
    \label{fig:fig1}
\end{figure}

The samples are grown on semi-insulating GaAs (100) substrates.  The lower part of the structure is composed of a 300~nm undoped GaAs buffer layer, followed by 2500~nm of Si-doped $n$-GaAs ($n= 3.6\times 10^{16}$~cm$^{-3}$), which forms the channel of the device.  The junction region consists of a 15~nm $n\rightarrow n^{+}$ GaAs transition layer followed by 15~nm $n^{+}$ (5 $\times$ 10$^{18}$~cm$^{-3}$) GaAs \cite{Hanbicki:2003}.  The Fe film, 5~nm thick, is then deposited epitaxially at $\sim 0$~$^\circ$C, followed by a 2 nm Al capping layer.  A Schottky tunnel barrier is formed by the Fe and the highly doped GaAs layer.  Samples with other channel doping levels were also prepared, as well as a control sample with the Fe film replaced with Al.  The heterostructures are processed into Hall bars in which the heavily doped GaAs layer at the top of the structure is removed everywhere except underneath the Fe contacts.  A photomicrograph of a typical device is shown in Fig. \ref{fig:fig1}(a).  The six Fe contacts are labelled {\it a - e}.   The  source and drain contacts {\it a} and {\it b} are separated by 150 $\mu$m, which is much longer than the spin diffusion length ($\sim 10$~$\mu$m at 10~K) and spin drift length ($\sim 30$~$\mu$m at 10 K and 1.0 mA).  The easy magnetization axis of each $40\times 100$~$\mu$m Fe contact is parallel to the channel, which is along the GaAs [011] direction.

The calculated band diagram of the  Fe/GaAs Schottky contacts is shown in Fig. \ref{fig:fig1}(b) for three bias voltages covering the range of our experiment.  Experimental I-V curves at $T=10$~K are shown in the inset.  In addition to the voltage $V_{ab}$ measured from the source (reverse-biased) to the drain (forward-biased), we also measure the other voltages indicated in Fig.~\ref{fig:fig1}(a).  This approach allows us to determine the voltage drops at the source and drain contacts independently.

The lateral geometry allows us to image the spin polarization in the GaAs channel near the source and drain contacts using the magneto-optical Kerr effect.  We follow the approach of Ref.~\onlinecite{Crooker:2005} to verify that an electron spin polarization exists at both the source and drain electrodes in the presence of a current.  The polarization near the source electrode is due to spin injection through the Schottky barrier \cite{Zhu:2001,Hanbicki:2003,Adelmann:2005},  while the spin accumulation at the drain electrode is due to the spin-dependent reflectivity of the barrier under forward bias \cite{Epstein:2003,Stephens:2004,Ciuti}.  In either case, the polarization in the semiconductor is oriented antiparallel to the magnetization of the Fe contact \cite{Crooker:2005} and is suppressed by applying a magnetic field of $\sim$100~Oe perpendicular to the magnetization axis, demonstrating the existence of a Hanle effect.

\begin{figure}
    \includegraphics*[width=6.5cm]{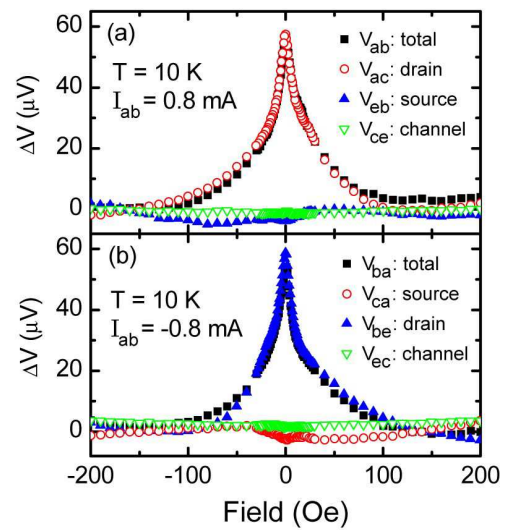}
    \caption{(color on-line) (a) Voltage measurements for a current $I = 0.8$~mA flowing from $a$ to $b$ as a function of magnetic field.  The source-drain voltage $V_{ab}\approx 0.5$~V.  An offset is subtracted from each curve, and a slope due to the Hall effect has been subtracted from $V_{ac}$ and $V_{eb}$.   The  contacts are labelled in Fig.~\ref{fig:fig1}(a). (b) Measurements for  $I = 0.8$~mA flowing from $b$ to $a$. For either current direction, the peak at zero field appears only for measurements that include the drain contact.}
    \label{fig:fig2}
\end{figure}
Given the existence of non-equilibrium spin populations at the source and drain, we consider the electrical transport properties of the device when a small magnetic field ($H\ll 4\pi M$) is applied perpendicular to the plane of the structure.  In this geometry, the magnetization of each electrode remains fixed during the measurement.   Fig.~\ref{fig:fig2}(a) shows voltage measurements at 10~K  for a 0.8~mA bias current flowing from contact $a$ to $b$, so that electrons are flowing from $b$ (source) to $a$ (drain).  The total source-drain voltage drop $V_{ab}$ is approximately 0.5~V.   A constant offset $V_0$ is subtracted from each curve, and a linear slope due to the Hall effect has also been removed from the data for $V_{ac}$ and $V_{ce}$.  The important feature of Fig.~\ref{fig:fig2}(a) is the peak at zero field in the total voltage $V_{ab}$ as well as the drain voltage $V_{ac}$. These two measurements nearly coincide and have an amplitude $\Delta V \sim 60$ ~$\mu$V and full-width at half maximum (FWHM) $\sim$~40~Oe.  In contrast, no peak is observed in the source and channel voltages $V_{eb}$ and $V_{ce}$.   When  the current direction in the channel is reversed, as for the data shown in Fig.~\ref{fig:fig2}(b), the peak is observed only for the voltage measurements that include contact $b$, which is now the drain.   The amplitude and FWHM of the peak are the same as in Fig.~\ref{fig:fig2}(a).  

The asymmetry between source and drain shown in Fig.~\ref{fig:fig2} has been observed in every Fe/GaAs device of this design that we have studied over the channel doping range from $2\times 10^{16}$~cm$^{-3}$ to $1.5\times 10^{17}$~cm$^{-3}$.  No peak is observed in devices in which the Fe has been replaced by Al.  Since the the electrode magnetization remains fixed, stray magnetic fields or anisotropic magnetoresistance can be excluded as possible origins of the effect.  This conclusion is reinforced by the fact that the additional voltage drop is present only at the drain contact.

The voltage peak at zero field observed at the drain contact is a signature of spin accumulation at the ferromagnet-semiconductor interface.   The effective resistance of the Schottky barrier is higher when a spin accumulation in the semiconductor is present.  This is consistent with the model of Ciuti {\it et al.}\cite{Ciuti}, in which the barrier is less transmissive for the spin state that accumulates in the semiconductor. The accumulation is suppressed by precession of the spin in a transverse field, leading to a decrease in the voltage.  This explanation is also consistent with information inferred from the optical detection experiments \cite{Crooker:2005}, with which a more detailed comparison will be made below.  As noted above, however, a non-equilibrium spin polarization is detected optically at {\it both} the source and drain electrodes, but the zero-field peak in the voltage is observed only at the drain contact.   This reflects the fact that the Fe/GaAs interface acts as a tunnel barrier.   At the reverse-biased source contact, the injected electrons tunnel from filled states in the ferromagnet into empty states in the semiconductor. These states are at an energy $eV_s\gg k_B T$ above the quasi-Fermi level for each spin, where $V_s$ is the voltage drop across the source contact.   The fractional occupancy of these states is therefore small, and so the number of available states for each spin is essentially the same.  As a result, the tunnelling resistance under reverse bias does not depend on the spin polarization in the semiconductor.  The situation at the forward-biased drain contact is very different, since both the filled states in the semiconductor and the empty states in the ferromagnet have significant polarizations.  In this case, the tunnelling current should be sensitive to the spin polarization in the semiconductor, as is observed experimentally.

\begin{figure}
    \includegraphics*[width=6.0cm]{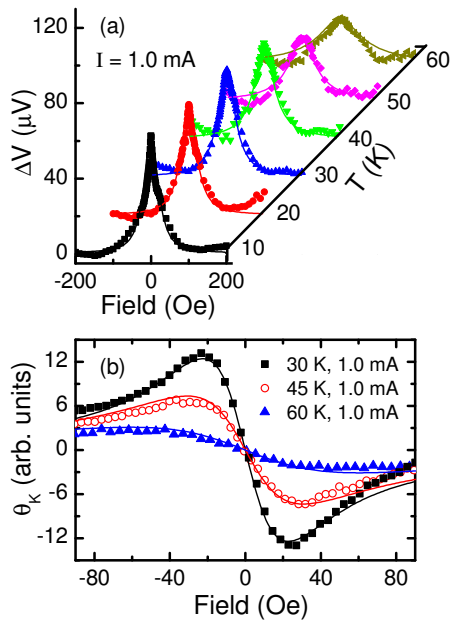}
    \caption{(color on-line) (a) The source-drain voltage $\Delta V_{ab}$ is shown as a function of magnetic field and temperature. Solid lines are the results of the modelling explained in the text.  (b)  The Kerr rotation $\theta_K$ (proportional to $S_z$), measured near the source,  is shown as a function of magnetic field for three temperatures.  The solid curves are calculated using the same parameters used to fit the transport data at each temperature (parameters at 45~K are obtained by interpolation).}
    \label{fig:fig3}
\end{figure}
Measurements of $\Delta V_{ab}$ for different temperatures at a fixed bias current of 1.0~mA are shown in Fig.~\ref{fig:fig3}(a).  To model these data, we adopt an approach used previously for spin-dependent transport in metals \cite {Johnson:1985,Jedema:2002}, but in this case the spins are generated and detected at a single ferromagnetic contact (the drain).   We assume a mechanism in which the electrons in GaAs are polarized with an orientation along the magnetization axis $\hat{x}$ by scattering from the interface at a point $x_1$ \cite{Ciuti}.  This process creates spin polarization at a rate $S_0$ per unit contact length.  The spins then drift and diffuse under the contact before being detected at some time $t$ at another position $x_2$.  As the spins diffuse, they also precess about the applied field $B$ and decay with a relaxation time $\tau_s$.  The $x$-component of the steady-state spin polarization $S_x(x_1,x_2,B)$ at $x_2$ due to spin generated at $x_1$ is obtained by integrating the Green's function solution of the drift-diffusion equation (including the spin relaxation and precession terms) over time, so that \cite{Johnson:1985,Jedema:2002,Stephens:2004,Crooker:2005}
\begin{eqnarray}
S_{x}(x_1, x_2, B) & = &\int_{0}^{\infty}\frac{S_{0}}{\sqrt{4 \pi Dt}} e^{-(x_2-x_1+v_{d}t)^{2}/4Dt- t/\tau_s } \nonumber \\
& &\times\cos(\frac{g\mu_{B}B}{\hbar}t)dt,
\label{eq:driftdiff}
\end{eqnarray}
where $D$ is the diffusion constant, $v_d$ is the drift velocity, $g=-0.44$ is the $g$-factor for electrons in GaAs, and $\mu_B$ is the Bohr magneton.  The average spin polarization $S_x(B)$ is determined by integrating $S_x(x_1,x_2,B)$ over $x_1$ and $x_2$, which vary from 0 to $L,$ where $L=40$~$\mu$m is the length of the drain contact.

With the exception of $S_0$, the parameters in Eq.~\ref{eq:driftdiff} are determined at each temperature by independent measurements.  The drift velocity $v_d$ is calculated from the current using the carrier density determined from Hall measurements.  The spin diffusion constant $D$ is determined from the measured Hall mobility, accounting for the temperature dependence of the Fermi distribution function at this density\cite{Flatte:2000}.  The spin lifetime $\tau_s$ is determined by the optical Hanle effect as measured by Kerr rotation. We assume that the voltage drop at the drain is proportional to $S_x(B)$ and fit the data of Fig.~\ref{fig:fig3}(a) to the expected form of $S_x(B)$ with $S_0$ (effectively the amplitude of the peak) as the only free parameter.

This procedure produces good agreement with the data at high temperatures, but the widths of the modelled peaks at low temperature are too small, with a discrepancy of approximately a factor of two at 10~K.  This reflects a greater sensitivity of the model at low temperatures to the time $L/v_d$ for the electrons to drift across the contact.  In practice, the current through the interface will be concentrated near the upstream edge of the contact and will be smallest near the downstream edge, with a gradient determined by the resistances of the channel and the interface, which are comparable [see inset of Fig.~\ref{fig:fig1}(b)].
We account for this by integrating over an effective length $L_e < L$ that is adjusted to fit one set of data (at 10 K) and then fixed for all other data sets.  The solid curves in Fig.~\ref{fig:fig3}(a) are calculated using $L_{e}=15$~$\mu$m.

An important aspect of the data that is predicted by the model is the increase in the FWHM of the peak as $T$ increases above 30~K.  This is due to the temperature dependence of $\tau_s$ \cite{Kikkawa:1998}, which decreases in this sample from 45~nsec at 10~K to 5~nsec at 60~K.  The peaks continue to broaden and decrease in amplitude above 60~K, but they can no longer be resolved above 80~K.

We can also show that the field dependence of the spin polarization that we measure electrically is the same as that measured by more traditional optical methods.  To demonstrate this, a small magnetic field is applied in the plane of the sample (perpendicular to the magnetization), and the spin polarization is measured using Kerr microscopy \cite{Crooker:2005}.  Electrically injected spins precess into the $z$-direction, leading to a Kerr rotation $\theta_K$ proportional to $S_z$.  Fig.~\ref{fig:fig3}(b) shows $\theta_K$  measured at the edge of the source contact at three different temperatures.  The solid curves are generated using the same parameters as for the transport data of Fig.~\ref{fig:fig3}(a) (except for the amplitude), with Eq.~\ref{eq:driftdiff} modified to account for the fact that $\theta_K$ is sensitive to $S_z$ rather than $S_x$.  In this case, we integrate only over the source coordinate because $\theta_K$ is measured at a single point.  The agreement with experiment is very good, providing extremely strong support for the interpretation of the transport measurements.

\begin{figure}
    \includegraphics*[width=6.0cm]{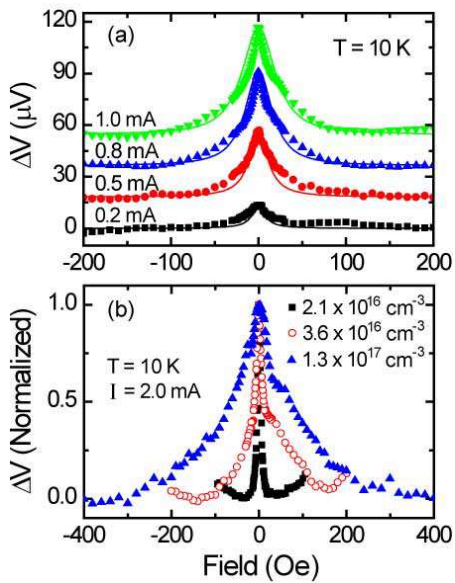}
    \caption{(color on-line) (a) The source-drain voltage $\Delta V_{ab}$ at $T=10$~K as a function of magnetic field for four different bias currents.  Each curve is offset for clarity.  The solid lines are the fitted curves using the model explained in the text. (b) $\Delta V_{ab}$ at $T=10$~K and $I=2.0$~mA for three $n$-type samples with the channel dopings indicated in the legend.}
    \label{fig:fig4}
\end{figure}

Besides $\tau_s$, the other timescales that can determine the FWHM of the voltage peak are $L_{e}^2/D$, $L_{e}/v_d$, and $4D/v_d^2$.  These have little effect at high temperature because they are all much larger than $\tau_s$.  At low temperatures, however, the two timescales that depend on $v_d$ become shorter than $\tau_s$ at the highest bias currents.  The source-drain voltage $V_{ab}$ is shown for several different bias currents at 10~K in Fig.~\ref{fig:fig4}(a).  The solid curves are generated from the model with $L_e=15$~$\mu$m. 
The amplitude, which is the only other parameter, is nearly proportional to the current.  The FWHM of the model curves also increases with current, due primarily to the fact that $L_{e}/v_d$ decreases from 130~nsec at 0.2~mA to 20~nsec at 1.0~mA.  The limiting width at small bias is set by spin relaxation ($\tau_s=45$~nsec at 10~K) and diffusion. 

The width also increases with bias current in the experimental data. At the highest currents, however, the peaks become sharper in the region $|B|< 10$~Oe.  This can be seen more prominently in Fig.~\ref{fig:fig4}(b), which shows $\Delta V_{ab}$ at a current of 2.0~mA for this sample ($n= 3.6\times 10^{16}$~cm$^{-3}$) and two others with different channel dopings.   The narrow features near zero field become stronger with decreasing doping and are characteristic of an enhancement of the effective magnetic field acting on the electrons due to dynamic polarization of nuclei \cite{DNP}.  This effect is regularly observed in optical pumping experiments and is also seen in optical measurements of the spin accumulation under forward bias \cite{Epstein:2003,Stephens:2004}.  Excluding very small fields, Fig.~\ref{fig:fig4}(b) shows an overall correlation between the FWHM and doping that is expected given the dependence of $\tau_s$ on carrier density \cite{Kikkawa:1998,Dzhioev:2002}. For example, the highest FWHM is observed for the sample with the largest channel doping ($1.3 \times 10^{17}$~cm$^{-3}$).

The measurements discussed here show that the electrically generated spin accumulation at a forward-biased $n$-GaAs/Fe Schottky contact can also be detected electrically.   A quantitative comparison is made with the established approach of optical detection.  The results demonstrate that it is possible to determine important parameters of a semiconductor spin transport device using electrical measurements alone.

This work was supported by the DARPA SpinS and Los Alamos LDRD programs, the NSF MRSEC program under DMR 02-12032, ONR, and the NSF NNIN program through the Minnesota Nanofabrication Center.

\end{document}